\documentclass[10pt,a4paper]{article}
\usepackage{amsmath}
\usepackage{amssymb}
\usepackage{amsfonts}
\usepackage{amstext}
\usepackage{amsbsy}
\usepackage[mathscr]{eucal}
\usepackage{graphicx}
\usepackage{color}
\usepackage[all]{xy}
\newcommand{\beq}{\begin{equation}}
\newcommand{\eeq}{\end{equation}}
\newcommand{\beqa}{\begin{eqnarray}}
\newcommand{\eeqa}{\end{eqnarray}}
\newcommand{\ket}[1]{| #1 \rangle}

\makeindex
\title{\Large\textbf{Quantum entangled state, resolution  of  conifold singularity, and toric variety}}

\author{\textit{ Hoshang Heydari}\\
        \small\textit{Physics Department, Stockholm university 10691 Stockholm Sweden}\\
\\\small\textit{Email: hoshang@fysik.su.se}}

\date{}
\begin{document}

\maketitle \thispagestyle{empty}

\maketitle
\begin{abstract}
We investigate the geometrical and combinatorial structures of multipartite quantum systems based on conifold and toric variety. In particular, we study the relations between resolution of conifold, toric variety, a separable state, and a quantum entangled state for bipartite and multi-qubit states. For example we show that the resolved or deformed conifold is equivalent with the space of a pure entangled  two-qubit state. We also generalize this result into multi-qubit states.
\end{abstract}


\section{Introduction}
Pure quantum states are usually defined on complex Hilbert spaces which are very complicated to visualize. The simplest case, namely, the space of a single qubit state can be visualized with Block or Riemann sphere. Beyond that there have been little progresses to visualize quantum state. Recently, we have established a relation between quantum states and toric varieties. Based on such a construction or mapping it is possible to visualize the complex Hilbert space by lattice polytop.

In algebraic
geometry \cite{Griff78}, a conifold is a generalization of the notion of a
manifold. But, a conifold can contain conical
singularities, e.g., points whose neighborhood look like a cone
with a certain base. The base is usually a five-dimensional
manifold. However, the base of a  complex conifold is a product of one dimensional complex projective space. Conifold are interesting space in string theory, e.g., in
the process of compactification of Calabi-Yau manifolds. A
Calabi-Yau manifold is a compact K\"{a}hler manifold with a
vanishing first Chern class. A Calabi-Yau manifold can also be
defined as a compact Ricci-flat K\"{a}hler manifold.

During recent decade toric varieties have been constructed in different
contexts in mathematics \cite{Ewald,GKZ,Fulton}. A toric variety $\mathbb{X}$ is a complex
variety that contains an algebraic torus $T=(\mathbb{C}^{*})^{n}$
as a dense open set and with action of $T$ on $\mathbb{X}$  whose restriction to
 $T\subset\mathbb{X}$ is the usual multiplication on $T$.

In this paper, we establish relations between toric varieties and space of entangled states of bipartite and multipartite quantum systems. In particular, we discuss resolving the singularity and deformation of conifold and toric variety of the conifold. We show that by removing the singularity of conifold variety we get a space which not anymore toric variety but is the space of an entangled two-qubit state. We also investigate the combinatorial structure of multi-qubit systems based on deformation of each faces of cube (hypercube) which is equivalent to deformation of conifold variety.
Through this paper we will use  the following notation
\begin{equation}\label{qubit}
\ket{\Psi}=\sum^{1}_{x_{m}=0}\sum^{1}_{x_{m-1}=0}\cdots
\sum^{1}_{
x_{1}=0}\alpha_{x_{m}x_{m-1}\cdots x_{1}}\ket{x_{m}x_{m-1}\cdots
x_{1}},
\end{equation}
 with $\ket{x_{m}x_{m-1}\cdots
x_{1}}=\ket{x_{m}}\otimes\ket{x_{m-1}}\otimes\cdots\otimes\ket{x_{1}}\in
\mathcal{H}_{\mathcal{Q}}=\mathcal{H}_{\mathcal{Q}_{1}}\otimes
\mathcal{H}_{\mathcal{Q}_{2}}\otimes\cdots\otimes\mathcal{H}_{\mathcal{Q}_{m}}
$ for a pure multi-qubit state.

\section{Conifold}
In this section we will give a short review of conifold.
Let $\mathbb{C}$ be a complex algebraic field. Then, an affine
$n$-space over $\mathbb{C}$ denoted $\mathbb{C}^{n}$ is the set of
all $n$-tuples of elements of $\mathbb{C}$. An element
$P\in\mathbb{C}^{n}$ is called a point of $\mathbb{C}^{n}$ and if
$P=(a_{1},a_{2},\ldots,a_{n})$ with $a_{j}\in\mathbb{C}$, then
$a_{j}$ is called the coordinates of $P$.
 A complex projective space $\mathbb{P}_{\mathbb{C}}^{n}$ is
defined to be the set of lines through the origin in
$\mathbb{C}^{n+1}$, that is,
\begin{equation}
\mathbb{P}_{\mathbb{C}}^{n}=\frac{\mathbb{C}^{n+1}-\{0\}}{
(x_{0},\ldots,x_{n})\sim(y_{0},\ldots,y_{n})},~\lambda\in
\mathbb{C}-0,~y_{i}=\lambda x_{i}
\end{equation}
 for all $0\leq i\leq n $.
An
example of real (complex) affine variety is conifold which is
defined by
\begin{equation}
\mathcal{V}_{\mathbb{C}}(z)=\{(z_{1},z_{2},z_{3},z_{4})
\in\mathbb{C}^{4}: \sum^{4}_{i=1}z^{2}_{i}=0\}.
\end{equation}
and conifold as a real affine variety is define by
\begin{equation}
\mathcal{V}_{\mathbb{R}}(f_{1},f_{2})=\{(x_{1},\ldots,x_{4},y_{1},\ldots,y_{4})\in\mathbb{R}^{8}:
\sum^{4}_{i=1}x^{2}_{i}=\sum^{4}_{j=1}y^{2}_{j},\sum^{4}_{i=1}x_{i}y_{i}=0
\}.
\end{equation}
where $f_{1}=\sum^{4}_{i=1}(x^{2}_{i}-y^{2}_{i})$ and
$f_{2}=\sum^{4}_{i=1}x_{i}y_{i}$. This can be seen by defining
$z=x+iy$ and identifying imaginary and real part of equation
$\sum^{4}_{i=1}z^{2}_{i}=0$.  As a real space, the conifold
is cone in $\mathbb{R}^{8}$ with top the origin and base space the
compact manifold $\mathbb{S}^{2}\times\mathbb{S}^{3}$.
  One can reformulate this relation in
term of a theorem. The conifold $
\mathcal{V}_{\mathbb{C}}(\sum^{4}_{i=1}z^{2}_{i}) $ is the complex
cone over the Segre variety $\mathbb{CP}^{1}
  \times\mathbb{CP}^{1}\longrightarrow\mathbb{CP}^{3}$. To see this let us make a complex linear
  change of coordinate
 \begin{equation}
 \left(
   \begin{array}{cc}
    \alpha^{'}_{00} & \alpha^{'}_{01} \\
     \alpha^{'}_{10} & \alpha^{'}_{11}\\
   \end{array}
 \right)\longrightarrow \left(
   \begin{array}{cc}
    z_{1}+iz_{2} & -z_{4}+iz_{3} \\
    z_{4}+iz_{3} & z_{1}-iz_{2}\\
   \end{array}
 \right).
 \end{equation}
    Thus after this linear
  coordinate transformation we have
  \begin{equation}\label{Conifold}
    \mathcal{V}_{\mathbb{C}}(\alpha^{'}_{00}\alpha^{'}_{11}-\alpha^{'}_{01}\alpha^{'}_{10})
    =\mathcal{V}_{\mathbb{C}}(\sum^{4}_{i=1}z^{2}_{i})\subset\mathbb{C}^{4}.
\end{equation}
Thus we can think of conifold as a complex cone over $\mathbb{CP}^{1}
  \times\mathbb{CP}^{1}$ see Figure 1.
  \begin{figure}\label{fig1a}
\begin{center}
\includegraphics[scale=0.450]{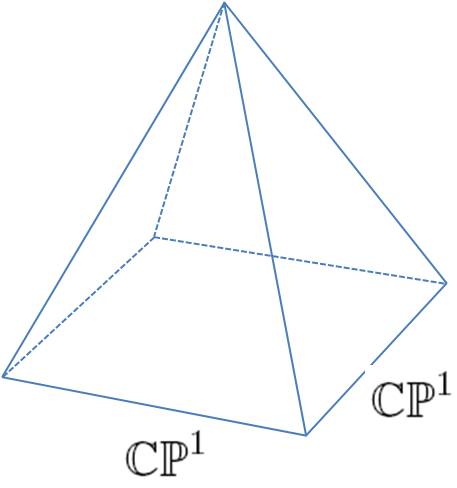}
\end{center}
\caption{Complex cone over $\mathbb{CP}^{1}
  \times\mathbb{CP}^{1}$.}
\end{figure}
We will comeback to this result in section \ref{coifoldsec} where
we establish a relation between these varieties,  two-qubit state, resolution of singulary, and deformation theory.  We can also define a metric on conifold as
$dS^{2}_{6}=d r^{2}+r^{2}d S^{2}_{T^{1,1}}$, where
\begin{equation}
d S^{2}_{T^{1,1}}=\frac{1}{9}\left(d\psi +\sum^{2}_{i=1}\cos
\theta_{i}d\phi_{i}\right)^{2}+\frac{1}{6}\sum^{2}_{i=1}\left(d\phi^{2}_{i}
+\sin^{2} \theta_{i}d\phi^{2}_{i}\right)^{2},
\end{equation}
is the metric on the Einstein manifold $T^{1,1}=\frac{SU(2)\times
SU(2)}{U(1)}$, with $U(1)$ being a diagonal subgroup of the
maximal torus of $SU(2)\times SU(2)$. Moreover, $T^{1,1}$ is a
$U(1)$ bundle over $\mathbb{S}^{2}\times\mathbb{S}^{2}$, where
$0\leq \psi\leq 4$ is an angular coordinate and
$(\theta_{i},\phi_{i})$ for all $i=1,2$ parameterize the two
$\mathbb{S}^{2}$, see Ref. \cite{Kleb,Morr,Hosh6}.

\section{Toric varieties}
\label{sec:2}
The construction of toric varieties usually are based on two different branches of mathematics, namely, combinatorial geometry and algebraic geometry. Here, we will review the   basic notations and structures of toric varieties \cite{Ewald,GKZ,Fulton}.

A general toric variety is an irreducible variety $\mathbb{X}$ that satisfies the following conditions. First of all $(\mathbb{C}^{*})^{n}$ is a Zariski open subset of $\mathbb{X}$ and the action of
 $(\mathbb{C}^{*})^{n}$  on itself can extend to an action of  $(\mathbb{C}^{*})^{n}$  on the variety
 $\mathbb{X}$. As an example we will show that the complex projective space $\mathbb{P}^{n}$ is a toric variety. If $z_{0},z_{1},
 \ldots,z_{n}$ are homogeneous coordinate of  $\mathbb{P}^{n}$. Then, the map
 $(\mathbb{C}^{*})^{n}\longrightarrow\mathbb{P}^{n}$ is defined by $(t_{1},t_{2},
 \ldots,t_{n})\mapsto(1,t_{1},
 \ldots,t_{n})$  and we have
\begin{equation}
 (t_{1},t_{2},
 \ldots,t_{n})\cdot (a_{0},a_{1},
 \ldots,a_{n})=(a_{0},t_{1}a_{1},
 \ldots,t_{n}a_{n})
\end{equation} which proof our claim that $\mathbb{P}^{n}$ is a toric variety.
 We can also define toric varieties with combinatorial information such as polytope and fan.
 But first we will give a short introduction to the basic of combinatorial geometry which is important in definition of toric varieties. Let $S\subset \mathbb{R}^{n}$ be finite subset, then a convex polyhedral cone is defined by
\begin{equation}
 \sigma=\mathrm{Cone}(S)=\left\{\sum_{v\in S}\lambda_{v}v|\lambda_{v}\geq0\right\}.
\end{equation}
In this case $\sigma$ is generated by $S$.  In a similar way we define  a polytope by
\begin{equation}
 P=\mathrm{Conv}(S)=\left\{\sum_{v\in S}\lambda_{v}v|\lambda_{v}\geq0, \sum_{v\in S}\lambda_{v}=1\right\}.
\end{equation}
We also could say that $P$ is convex hull of $S$. A convex polyhedral cone is called simplicial if it is generated by linearly independent set. Now, let $\sigma\subset \mathbb{R}^{n}$ be a convex polyhedarl cone and $\langle u,v\rangle$ be a natural pairing between $u\in \mathbb{R}^{n}$ and $v\in\mathbb{R}^{n}$. Then, the dual cone of the $\sigma$ is define by
\begin{equation}
 \sigma^{\wedge}=\left\{u\in \mathbb{R}^{n*}|\langle u,v\rangle\geq0~\forall~v\in\sigma\right\},
,
\end{equation}
where $\mathbb{R}^{n*}$ is dual of $\mathbb{R}^{n}$. We also define the polar of $\sigma$ as
\begin{equation}
 \sigma^{\circ}=\left\{u\in \mathbb{R}^{n*}|\langle u,v\rangle\geq-1~\forall~v\in\sigma\right\}.
\end{equation}
 We call a convex polyhedral cone strongly convex if $\sigma\cap(-\sigma)=\{0\}$.

Next we will define rational polyhedral cones. A free Abelian group of finite rank is called a lattice, e.g., $N\simeq\mathbb{Z}^{n}$. The dual of a lattice $N$ is defined by
$M=\mathrm{Hom}_{\mathbb{Z}}(N,\mathbb{Z})$ which has rank $n$. We also define a vector space and its dual by $N_{\mathbb{R}}=N\otimes_{\mathbb{Z}}\mathbb{R}\simeq \mathbb{R}^{n}$ and $M_{\mathbb{R}}=M\otimes_{\mathbb{Z}}\mathbb{R}\simeq \mathbb{R}^{n*}$ respectively.
 Moreover, if $\sigma=\mathrm{Cone}(S)$ for some finite set $S\subset N$, then $\sigma\subset N_{\mathbb{R}}$ is a rational polydehral cone. Furthermore, if  $\sigma\subset N_{\mathbb{R}}$ is a rational polyhedral cone, then   $S_{\sigma}=\sigma^{\wedge}\cap M$ is a semigroup under addition with $0\in S_{\sigma}$ as additive identity which is finitely generated by Gordan's lemma \cite{Ewald}.

Here we will define a fan which is important in the construction of toric varieties. Let $\Sigma\subset N_{\mathbb{R}}$ be a finite non-empty set of strongly convex rational polyhedral cones. Then $\Sigma$ is called a fan if each face of a cone in $\Sigma$  belongs to $\Sigma$ and the intersection of any two cones in $\Sigma$  is a face of each.

 Now, we can obtain the coordinate ring of a variety by associating to the semigroup $S$ a finitely generated commutative $\mathbb{C}$-algebra without nilpotent as follows. We associate  to an arbitrary additive semigroup its semigroup algebra $\mathbb{C}[S]$ which as a vector space has the set $S$ as basis. The elements of $\mathbb{C}[S]$ are linear combinations
 $\sum_{u\in S}a_{u}\chi^{u}$ and the product in $\mathbb{C}[S]$ is determined by the addition in $S$ using  $\chi^{u}\chi^{u^{'}}=\chi^{u+u^{'}}$ which is called the exponential rule. Moreover, a set of semigroup generators $\{u_{i}: i\in I\}$ for $S$ gives algebra generators $\{\chi^{u_{i}}: i\in I\}$ for  $\mathbb{C}[S]$.

 Now, let $\sigma\subset N_{\mathbb{R}}$ be a strongly convex rational polyhedral cone and  $A_{\sigma}=\mathbb{C}[S_{\sigma}]$ be an algebra which is a normal domain. Then,
\begin{equation}
\mathbb{X}_{\sigma}=\mathrm{Spec}(\mathbb{C}[S_{\sigma}])=\mathrm{Spec}(A_{\sigma})
\end{equation}
is called a affine toric variety.   Next we need to define Laurent polynomials and monomial algebras. But first we observe that the dual cone $\sigma^{\vee}$ of the zero cone $\{0\}\subset N_{\mathbb{R}}$ is all of $ M_{\mathbb{R}}$ and the associated semigroup $S_{\sigma}$ is the group $M\simeq \mathbb{Z}^{n}$. Moreover, let $(e_{1},e_{2},\ldots,e_{n})$ be a basis of $N$ and
$(e^{*}_{1},e^{*}_{2},\ldots,e^{*}_{n})$ be its dual basis for $M$. Then, the elements $\pm e^{*}_{1},\pm e^{*}_{2},\ldots,\pm e^{*}_{n}$ generate $M$ as semigroup. The algebra of Laurent polynomials is defined by
\begin{equation}
\mathbb{C}[z,z^{-1}]=\mathbb{C}[z_{1},z^{-1}_{1},\ldots,z_{n},z^{-1}_{n}],
\end{equation}
where $z_{i}=\chi^{e^{*}_{i}}$. The terms  of the form $\lambda \cdot z^{\beta}=\lambda z^{\beta_{1}}_{1}z^{\beta_{2}}_{2}\cdots z^{\beta_{n}}_{n}$ for $\beta=(\beta_{1},\beta_{2},\ldots,\beta_{n})\in \mathbb{Z}$ and $\lambda\in \mathbb{C}^{*}$ are called Laurent monomials. A ring $R$ of Laurent polynomials is called a monomial algebra if it is a $\mathbb{C}$-algebra generated by Laurent monomials. Moreover, for a lattice cone $\sigma$, the ring
$R_{\sigma}=\{f\in \mathbb{C}[z,z^{-1}]:\mathrm{supp}(f)\subset \sigma\}
$
is a finitely generated monomial algebra, where the support of a Laurent polynomial $f=\sum\lambda_{i}z^{i}$ is defined by
$\mathrm{supp}(f)=\{i\in \mathbb{Z}^{n}:\lambda_{i}\neq0\}.
$ Now, for a lattice cone $\sigma$ we can define an affine toric variety to be the maximal spectrum $\mathbb{X}_{\sigma}=\mathrm{Spec}R_{\sigma}$.  A toric variety
$\mathbb{X}_{\Sigma}$ associated to a fan $\Sigma$ is the result of gluing affine varieties
$\mathbb{X}_{\sigma}=\mathrm{Spec}R_{\sigma}$ for all $\sigma\in \Sigma$ by identifying $\mathbb{X}_{\sigma}$ with the corresponding Zariski open subset in $\mathbb{X}_{\sigma^{'}}$ if
$\sigma$ is a face of $\sigma^{'}$. That is,
first we take the disjoint union of all affine toric varieties $\mathbb{X}_{\sigma}$ corresponding to the cones of $\Sigma$.
Then by gluing all these affine toric varieties together we get $\mathbb{X}_{\Sigma}$.
A affine toric variety $\mathbb{X}_{\sigma}$ is non-singular if and only if the normal polytope has a
unit volume.

\section{Conifold and resolution of toric singulrity for two-qubits}\label{coifoldsec}

In this section we study the simplicial decomposition of affine toric variety. For two qubits this simplicial decomposition coincides with desingularizing a conifold \cite{Closset}. We also show that resolved conifold is  space of an entangles two-qubit state.
For a pairs of qubits  $\ket{\Psi}=\sum^{1}_{x_{2}=0}\sum^{1}_{x_{1}=0}
\alpha_{x_{2}x_{1}} \ket{x_{2}x_{1}}$ we can also construct following simplex. For this two qubit state the separable state is given by the Segre embedding of $\mathbb{CP}^{1}\times\mathbb{CP}^{1}=
\{((\alpha^{1}_{0},\alpha^{1}_{1}),(\alpha^{2}_{0},\alpha^{2}_{1})): (\alpha^{1}_{0},\alpha^{1}_{1})\neq0,~(\alpha^{2}_{0},\alpha^{2}_{1})\neq0\}$. Let $z_{1}=\alpha^{1}_{1}(\alpha^{1}_{0})^{-1}$ and $z_{2}=\alpha^{2}_{1}(\alpha^{2}_{0})^{-1}$. Then we can cover $\mathbb{CP}^{1}\times\mathbb{CP}^{1}$ by four charts
\begin{equation}
\mathbb{X}_{\check{\Delta}_{1}}=\{(z_{1},z_{2})\},
~\mathbb{X}_{\check{\Delta}_{2}}=\{(z^{-1}_{1},z_{2})\},~
\mathbb{X}_{\check{\Delta}_{3}}=\{(z_{1},z^{-1}_{2})\},~
\mathbb{X}_{\check{\Delta}_{4}}=\{(z^{-1}_{1},z^{-1}_{2})\},
\end{equation}
 The fan $\Sigma$ for $\mathbb{CP}^{1}\times\mathbb{CP}^{1}$ has edges spanned by $(1,0),(0,1),(-1,0),(0,-1)$. Next we observe that the space $\mathbb{CP}^{1}\times\mathbb{CP}^{1}$ and the conifold have the same toric variety.
If we split the conifold into a fan which has two cones as shown in Figure 2. Then this process converts the conifold into a resolved conifold. The cones are three dimensional and the dual cones are  two copies of $\mathbb{C}^{3}$. The procedure of replacing an isolated singularity by a holomorphic cycle is called a resolution of the singularity. The resolved conifold has a Ricci-flat K\"{a}hler metric which was derived by Candelas and de la Ossa \cite{Candelas}
\begin{eqnarray}
\nonumber
ds^{2}_{res}&=& \widetilde{\varrho}^{'}d\widetilde{r}^{2}
+\frac{\widetilde{\varrho}^{'}}{4}\widetilde{r}^{2}(d\widetilde{\psi}
+\cos\widetilde{\theta}_{1}d\widetilde{\phi}_{1}
+\cos\widetilde{\theta}_{2}d\widetilde{\phi}_{2})^{2}\\\nonumber&+&
\frac{\widetilde{\varrho}}{4}(d\widetilde{\theta}^{2}_{1}
+\sin^{2}\widetilde{\theta}_{1}d\widetilde{\phi}^{2}_{1})+
\frac{\widetilde{\varrho}+4a^{2}}{4}(d\widetilde{\theta}^{2}_{2}
+\sin^{2}\widetilde{\theta}_{2}d\widetilde{\phi}^{2}_{2}),
\end{eqnarray}
where $\widetilde{\psi}=0\cdots4\pi$ is a $U(1)$ fiber over $S^{2}$,  $(\widetilde{\phi}_{i},\widetilde{\theta}_{i}),~i=1,2$ are  Euler angles on two sphere $S^{2}$, and $\widetilde{\varrho}=\widetilde{\varrho}(\widetilde{r})$  goes to zero as $\widetilde{r}\rightarrow0$.
Thus we can propose that the resolution of  the singularity of a toric variety is equivalent to the space of entangled states.
\begin{figure}\label{fig1}
\begin{center}
\includegraphics[scale=0.450]{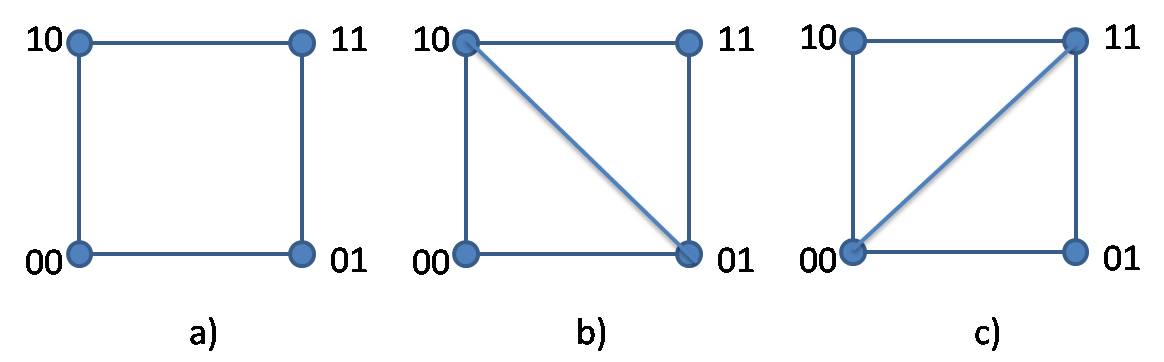}
\end{center}
\caption{Two-qubit system. a) toric polytope of a two-qubit systems. b) and c) two ways of removing the singularity of conifold.}
\end{figure}
We can also remove the singularity by deformation. The process of deformation modifies the complex structure manifolds or
algebraic varieties. Based on our discussion of conifold we know that this space is defined by
$\alpha_{00}\alpha_{11}-\alpha_{01}\alpha_{10}$. Now, if we rewrite this equation in the following form
\begin{equation}
\alpha_{00}\alpha_{11}-\Gamma\alpha_{01}\alpha_{10}+\Lambda\alpha_{10}=0,
\end{equation}
then the constant $\Gamma$ and $\Lambda$  can be absorbed in new definition of $\alpha_{10}$ such as
$\alpha^{'}_{10}=\Gamma\alpha_{10}-\Lambda$. Next let $T_{n}$ be the group of translations. Then an affine variety over complex field of dimension $n$ can be transformed using the following action
$GL(n,\mathbb{C})\ltimes T_{n}$. For a generic polynomial of degree two we have 15 possible parameters, but most of them can be removed with the action of $GL(4,\mathbb{C})\ltimes T_{4}$. However, we cannot remove the constant term with such transformation and we end up with the following variety
\begin{equation}
\alpha_{00}\alpha_{11}-\alpha_{01}\alpha_{10}=\Omega.
\end{equation}
which is called deformed conifold. This space is now non-singular, but it is not a toric variety since the deformation break one action of torus. Thus we also could proposed that  the deformed conifold is the space of an entangled pure two-qubit state. Moreover, if we take the absolute value of this equation that is $|\Omega|$, then this value is proportional to  concurrence which is a measure of entanglement for a pure two-qubit state, that is
\begin{equation}
|\alpha_{00}\alpha_{11}-\alpha_{01}\alpha_{10}|=|\Omega|=C(\Psi)/2.
\end{equation}
In general let $X$ be an algebraic variety, then the space of all complex deformations of $X$ is called the complex moduli space of $X$.

\section{Three-qubit states}
Next, we will discuss a three-qubit state $\ket{\Psi}=\sum^{1}_{x_{3},x_{2},x_{1}=0}
\alpha_{x_{3}x_{2}x_{1}} \ket{x_{3}x_{2}x_{1}}$.
For this  state the separable state is given by the Segre embedding of $\mathbb{CP}^{1}\times\mathbb{CP}^{1}\times\mathbb{CP}^{1}=
\{((\alpha^{1}_{0},\alpha^{1}_{1}),(\alpha^{2}_{0},\alpha^{2}_{1}),(\alpha^{3}_{0},\alpha^{3}_{1}))): (\alpha^{1}_{0},\alpha^{1}_{1})\neq0,~(\alpha^{2}_{0},\alpha^{2}_{1})\neq0
,~(\alpha^{3}_{0},\alpha^{3}_{1})\neq0\}$.
Now, for example, let $z_{1}=\alpha^{1}_{1}/\alpha^{1}_{0}$,
 $z_{2}=\alpha^{2}_{1}/\alpha^{2}_{0}$, and $z_{3}=\alpha^{3}_{1}/\alpha^{3}_{0}$.
  Then we can cover $\mathbb{CP}^{1}\times\mathbb{CP}^{1}\times\mathbb{CP}^{1}$ by eight charts
\begin{eqnarray}
\nonumber &&
\mathbb{X}_{\check{\Delta}_{1}}=\{(z_{1},z_{2},z_{3})\},
~\mathbb{X}_{\check{\Delta}_{2}}=\{(z^{-1}_{1},z_{2},z_{3})\},~
\mathbb{X}_{\check{\Delta}_{3}}=\{(z_{1},z^{-1}_{2},z_{3})\},~\\\nonumber&&
\mathbb{X}_{\check{\Delta}_{4}}=\{(z_{1},z_{2},z^{-1}_{3})\},
\mathbb{X}_{\check{\Delta}_{5}}=\{(z^{-1}_{1},z^{-1}_{2},z_{3})\},
~\mathbb{X}_{\check{\Delta}_{6}}=\{(z^{-1}_{1},z_{2},z^{-1}_{3})\},~\\\nonumber&&
\mathbb{X}_{\check{\Delta}_{7}}=\{(z_{1},z^{-1}_{2},z^{-1}_{3})\},~
\mathbb{X}_{\check{\Delta}_{8}}=\{(z^{-1}_{1},z^{-1}_{2},z^{-1}_{3})\},
\end{eqnarray}
 The fan $\Sigma$ for $\mathbb{CP}^{1}\times\mathbb{CP}^{1}\times\mathbb{CP}^{1}$ has edges spanned by $(\pm1,\pm1,\pm1)$.
Now, let $S=\mathbb{Z}^{3}$ and consider the polytope $\Delta$ centered at the origin with vertices $(\pm1,\pm1,\pm1)$. This gives the toric variety $\mathbb{X}_{\Delta}=\mathrm{Spec}\mathbb{C}[S_{\Delta}]$. To describe the fan of $\mathbb{X}_{\Delta}$, we observe that the polar $\Delta^{\circ}$ is the octahedron with vertices $\pm e_1,\pm e_2, \pm e_3$. Thus the normal fan is formed from the faces of the octahedron which gives a fan $\Sigma$ whose 3-dimensional cones are octants of $\mathbb{R}^{3}$. Thus this shows that the toric variety $\mathbb{X}_{\Sigma}=\mathbb{CP}^{1}\times\mathbb{CP}^{1}\times\mathbb{CP}^{1}$.

\begin{figure}\label{fig2}
\begin{center}
\includegraphics[scale=0.450]{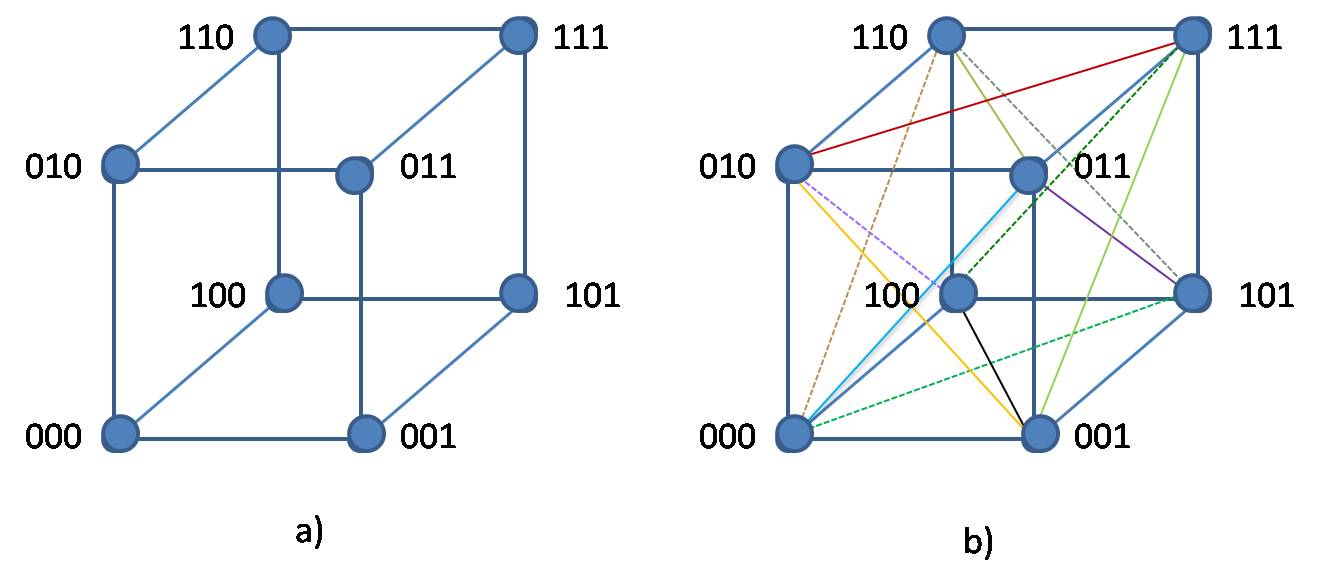}
\end{center}
\caption{Three-qubit systems. a) toric polytope of a separable three-qubit systems. b) resolved space of entangled state, where each diagonal line is equivalent to  the resolution of singularity of a conifold.   }
\end{figure}
In this case
 we split the faces of $3$-cube $E_{2,3}=2^{3-2}\frac{3(3-1)}{2}=6$ into  two cones see Figure 3. Then this process converts the $3$-cube into a nonsingular space which is not anymore toric variety.
Following the same procedure, we can also remove all singularities of toric variety of three-qubits by deformation.  Based on our discussion of conifold  we can write six equations describing the faces of 3-cube. Here we will analyze one face of this 3-cube, namely
\begin{eqnarray}
 \nonumber
 &&\alpha_{000}\alpha_{011}-\alpha_{001}\alpha_{010}=\alpha_{0}\otimes(\alpha_{00}
 \alpha_{11}-\alpha_{01}\alpha_{10})
\end{eqnarray}
Now, if we rewrite these equations e.g., in the following form
\begin{equation}
\alpha_{0}(\alpha_{00}\alpha_{11}-\Gamma\alpha_{01}\alpha_{10}+\Lambda\alpha_{10})=0,
\end{equation}
then the constant $\Gamma$ and $\Lambda$  can be absorbed in new definition of $\alpha_{10}$ such as
$\alpha^{'}_{10}=\Gamma\alpha_{10}-\Lambda$. At the end e.g., we  have the following variety
\begin{equation}
\alpha_{000}\alpha_{011}-\alpha_{001}\alpha_{010}=\Omega
\end{equation}
which is equivalent to the deformed conifold. If we  do this procedure for all faces of the 3-cube, then the whole space becomes non-singular, but it is not a toric variety anymore. Thus we also could proposed that  the deformed conifold is the space of an entangled pure three-qubit state.  There are other relations between toric variety and measures of quantum entanglement that can be seen from the toric structures of multipartite systems. For example three-tangle or 3-hyperdeterminant can be constructed from the toric variety.

\section{Multi-qubit states}
Next, we will discuss a multi-qubit state $\ket{\Psi}$ defined by equation (\ref{qubit}).
For this  state the separable state is given by the Segre embedding of $\mathbb{CP}^{1}\times\mathbb{CP}^{1}\times\cdots\times\mathbb{CP}^{1}=
\{((\alpha^{1}_{0},\alpha^{1}_{1}),(\alpha^{2}_{0},\alpha^{2}_{1}),
\ldots,(\alpha^{m}_{0},\alpha^{m}_{1}))): (\alpha^{1}_{0},\alpha^{1}_{1})\neq0,~(\alpha^{2}_{0},\alpha^{2}_{1})\neq0,\ldots,
,~(\alpha^{m}_{0},\alpha^{m}_{1})\neq0\}$.
Now, for example, let $z_{1}=\alpha^{1}_{1}/\alpha^{1}_{0}, z_{2}=\alpha^{2}_{1}/\alpha^{2}_{0},\ldots, z_{m}=\alpha^{m}_{1}/\alpha^{m}_{0}$.
  Then we can cover $\mathbb{CP}^{1}\times\mathbb{CP}^{1}\times\cdots\times\mathbb{CP}^{1}$ by $2^{m}$ charts
\begin{eqnarray}
\nonumber &&
\mathbb{X}_{\check{\Delta}_{1}}=\{(z_{1},z_{2},\ldots,z_{m})\},
~\mathbb{X}_{\check{\Delta}_{2}}=\{(z^{-1}_{1},z_{2},\ldots,z_{m})\}\\\nonumber&,\cdots,&
\mathbb{X}_{\check{\Delta}_{2^{m}-1}}=\{(z_{1},z^{-1}_{2},\ldots,z^{-1}_{m})\},~
\mathbb{X}_{\check{\Delta}_{2^{m}}}=\{(z^{-1}_{1},z^{-1}_{2},\ldots,z^{-1}_{m})\}
\end{eqnarray}
 The fan $\Sigma$ for $\mathbb{CP}^{1}\times\mathbb{CP}^{1}\times\cdots\times\mathbb{CP}^{1}$ has edges spanned by $(\overbrace{\pm1,\pm1,\ldots,\pm1}^{m})$.
Now, let $S=\mathbb{Z}^{m}$ and consider the polytope $\Delta$ centered at the origin with vertices $(\pm1,\pm1,\ldots,\pm1)$. This gives the toric variety $\mathbb{X}_{\Delta}=\mathrm{Spec}\mathbb{C}[S_{\Delta}]$. To describe the fan of $\mathbb{X}_{\Delta}$, we observe that the polar $\Delta^{\circ}$ is the octahedron with vertices $\pm e_1,\pm e_2, \ldots,\pm e_m$.  Thus this shows that the toric variety $\mathbb{X}_{\Sigma}=\mathbb{CP}^{1}\times\mathbb{CP}^{1}\times\cdots\times\mathbb{CP}^{1}$.
In this case
 we split the faces of $m$-cube
\begin{equation}
E_{2,m}=2^{m-2}\frac{m(m-1)}{2}
\end{equation}
 into  two cones. Then this process converts the $m$-cube into a nonsingular space which is not anymore toric variety.
Following the same procedure, we can also remove all singularities of toric variety of multi-qubits by deformation.  Based on our discussion of conifold  we can write six equations describing the faces of $m$-cube. For example for  one face (2-cube) of this $m$-cube, we ahve
\begin{equation}
\alpha_{00\cdots0}\alpha_{0\cdots011}-\alpha_{0\cdots01}\alpha_{0\cdots010}=\Omega
\end{equation}
which is equivalent to the deformed conifold, since e.g., we could have $\ket{\Psi}=\frac{1}{\sqrt{2}} (\ket{00\cdots000}+\ket{00\cdots011})=\frac{1}{\sqrt{2}} \ket{00\cdots0}\otimes(\ket{00}+\ket{11})$. If we  do this procedure for all faces of the m-cube, then the whole space becomes non-singular, but it is not a toric variety anymore. Thus we also could proposed that  this space  is the space of an entangled pure multi-qubit state.

In this paper we have investigated the geometrical and combinatorial structures of entangled multipartite systems. We have shown that by removing singularity of conifold or by deforming the conifold we obtain the space of a pure entangled two-qubit state. We have also generalized the construction into multipartite entangled systems.  The space of multipartite systems are difficult to visualize but the transformation from complex spaces to the combinatorial one makes this task much easier to realize. Hence our results give  new insight about multipartite systems  and also  a new way of representing quantum entangled bipartite and multipartite systems.

\begin{flushleft}
\textbf{Acknowledgments:} This  work was supported  by the Swedish Research Council (VR).
\end{flushleft}



\begin{thebibliography}{99}
\bibitem{Griff78} P. Griffiths and J. Harris, {\it Principle of
  algebraic geometry}, Wiley and Sons, New York, 1978.
\bibitem{Ewald}G. Ewald, {\it Combinatorial Convexity and Algebraic Geometry}, Springer, (1996).
\bibitem{GKZ} I. M. Gelfand,. M. M. Kapranov, and A.V. Zelevinsky, {\it Discriminants, resultants, and multidimensional determinants}, Birkhäuser, Boston, (1994),
\bibitem{Fulton} W. Fulton, {\it Introduction to Toric Varieties}, (1991).
\bibitem{Hosh5} H. Heydari and G.~Bj\"{o}rk, J. Phys. A: Math. Gen. {\bf 38}, 3203-3211 (2005).
\bibitem{Closset} C. Closset,  lectures given at the Modave Summer School in Mathematical Physics 2008, e-print 0901.3695v2.
\bibitem{Kleb} I. R. Klebanov, E. Witten, e-print hep-th/9807080.
\bibitem{Morr} D. R. Morrison and M. R. Plesser
, e-print hep-th/9810201.
\bibitem{Hosh6}H. Heydari, Quantum Information and Computation 6 (2006) 400-409.
\bibitem{Candelas} P. Candelas and X. C. de la Ossa, Nucl. Phys. B 342, 246.

\end{thebibliography}
\end{document}